\documentclass[twocolumn,superscriptaddress,showpacs]{revtex4-1} 
\usepackage{times,xspace} 
\usepackage{amsbsy,amssymb,amsmath,bm,bbold} 
\usepackage{graphicx,color,epsfig,rotate} 
\usepackage{fancyhdr} 
\usepackage{epstopdf}
\usepackage{float}
\usepackage{booktabs,makecell}
\usepackage{multirow}
\usepackage[table]{xcolor}
\usepackage[colorlinks=true,linkcolor=blue,citecolor=blue,urlcolor=blue]{hyperref}

\def\bbbc{{\mathchoice {\setbox0=\hbox{$\displaystyle\rm C$}\hbox{\hbox 
to0pt{\kern0.4\wd0\vrule height0.9\ht0\hss}\box0}} 
{\setbox0=\hbox{$\textstyle\rm C$}\hbox{\hbox 
to0pt{\kern0.4\wd0\vrule height0.9\ht0\hss}\box0}} 
{\setbox0=\hbox{$\scriptstyle\rm C$}\hbox{\hbox 
to0pt{\kern0.4\wd0\vrule height0.9\ht0\hss}\box0}} 
{\setbox0=\hbox{$\scriptscriptstyle\rm C$}\hbox{\hbox 
to0pt{\kern0.4\wd0\vrule height0.9\ht0\hss}\box0}}}}

\begin{document} 
\title{Nonlinearity Induced Anomalous Mode Collapsing and Non-chiral Asymmetric Mode Switching around multiple Exceptional Points}
\author{Sibnath Dey, Arnab Laha, and Somnath Ghosh}
\email{somiit@rediffmail.com}
\affiliation{Department of Physics, Indian Institute of Technology Jodhpur, Rajasthan-342037, India}

\begin{abstract} 
	
The dynamical encirclement around a second order exceptional point (EP) and corresponding chirality driven nonadiabatic modal dynamics have attracted enormous attention in the topological study of various non-Hermitian systems. However, dynamical encirclement around multiple second-order EPs in a multi-state system is yet to be explored. Here, exploiting an exclusive design of a planar gain-loss assisted three-mode supported optical waveguide with local Kerr-nonlinearity, we encounter multiple second-order EPs. Judiciously, choosing a specific parameter space by varying the unbalanced gain-loss profile, we encircle multiple EPs simultaneously, and explore the beam-dynamics toward corresponding chiral or non-chiral aspects of the device. While propagating through the designed waveguide, three coupled modes are collapsed into a specific dominating mode, owing to corresponding nonadiabatic corrections around multiple EPs. Even in the absence of chirality, here, the same amount of focusing and de-focusing type nonlinearity gives different dominating output, irrespective of the choice of inputs, for the same topological structure of the waveguide. This exclusive topologically robust compact scheme of nonlinearity induced asymmetric and non-chiral light dynamics should provide a promising opportunity to switch or retrieve a selective mode from a multi-mode signal in integrated devices.                 
     
\end{abstract} 
 
 
\maketitle %

Exhibiting Exceptional Points (EPs) is a nontrivial topological feature of open systems \cite{Kato} that have been substantially studied theoretically \cite{Sannino90,Mailybaev05} as well as experimentally \cite{Dembowski01} in almost all branches of non-Hermitian physics \cite{Miri19}. Especially in the topological photonics domain, using the optical gain-loss as non conservative ingredients \cite{Laha19,Ghosh16,Doppler16,Wong16,Ding15}, EPs have widely contributed to meet a wide range of benchmark applications like, asymmetric mode switching/conversion \cite{Ghosh16,Doppler16}, lasing and anti-lasing \cite{Wong16}, extreme enhancement in sensing \cite{Wiersig16}, optical isolation with enhanced nonreciprocity \cite{Thomas16}, etc. While an open system approaches an EP in parameter plane, the coupled eigenvalues coalesce in complex eigenvalue-plane; and simultaneously, the corresponding eigenvectors lose their identities and become self-orthogonal \cite{Kato,Sannino90}. A stroboscopic parametric encirclement enclosing a second order EP results in adiabatic flipping between a pair of coupled eigenmodes \cite{Dembowski01,Miri19,Laha19,Ghosh16} with an accumulation of Berry phase \cite{Mailybaev05}. In this context successive state-flipping in a multi-state system can be observed by encircling a higher-order EP \cite{Demange12,Pan19} or multiple second-order EPs \cite{Bhattacherjee19,Ryu12} in the system parameter space. A higher-order EP can be realized with coalescence of more than two coupled states \cite{Demange12,Pan19}, however, there are several investigations where similar unconventional physical effects associated with a higher-order EP have been realized by winding around multiple second-order EPs \cite{Bhattacherjee19,Ryu12,Ding16}.

Instead of stroboscopic encirclement around a second order EP, if we consider a time (or length-scale) dependent parametric variation to encircle the EP dynamically, then adiabaticity breaks down during state-evolutions \cite{Gilary13} in the sense that a clockwise and an anticlockwise parametric rotation results in different dominating state at the output, irrespective of the choice of the input state \cite{Ghosh16,Doppler16}. This chiral behavior due to the average effective loss difference between the coupled states during evolutions yields an asymmetric state-transfer phenomenon in practice \cite{Ghosh16,Doppler16}. Now, to consider a higher-order system, a natural question to be raised that whether the chiral property maintained for a second order EP connecting two eigenstates in the presence of other noninteracting states and again what would be the chiral aspect of the device if more than two states are mutually interacting in the vicinity multiple second-order EPs with proper parameter manipulation. In this context, the state-dynamics during the dynamical parametric encirclement around multiple second-order EPs or an higher-order EP in a multi-state system is yet to be explored. Beyond the already reported dual-mode systems \cite{Ghosh16,Doppler16}, it should be quite interesting and more compact from feasibility point-of-view in integrated devices, if it is possible to switch or retrieve a selective mode using few-mode or multi-mode systems.

Here, to address the highlighted issues, we investigate in a gain-loss assisted three-mode supported planar waveguide structure with local Kerr-nonlinearity. A particular topological structure of the waveguide with proper gain-loss variation has been judiciously chosen to modulate the interactions between the three supported modes. Initially, tuning the unbalanced gain-loss profile in the absence of nonlinearity, we encounter an EP between two coupled modes, keeping the third one unaffected, and dynamically encircling the identified EP, we study the chiral aspect of the device. In this letter, we establish the immutable chiral behavior of the device in the scene that depending on the encircling direction and corresponding EP-aided nonadiabatic corrections, a specific dominating mode from the pair of coupled modes survives, in simultaneous presence of the noninteracting mode. Now, with the onset of the nonlinearity in the optical medium, the previously unaffected mode is supposed to interact with the rest of the coupled modes. Simultaneously, varying the gain-loss profile, we encounter multiple second-order EPs to connect three coupled modes analytically. In this context, nonchiral behavior of three interacting eigenstates around multiple second-order EPs or a higher order EP was predicted analytically in a previous work \cite{Heiss08}. Now, simultaneously enclosing at least two EPs inside the dynamical parametric loop, we study the dynamics of the coupled eigenmodes. Here, we have shown that regardless of the choice of input mode, all of them are collapsed into a specific dominating mode, and most importantly, the chirality of the device is destroyed. An analytical model to describe this anomalous mode-collapsing phenomenon has been developed. Now, we exclusively investigate and report the influence of nonlinearity on the nonadiabatic mode-conversions around multiple EPs and establish that even in the absence of chirality, the same amount of focusing and de-focusing nonlinearity will lead to different dominating mode at the output.    
\begin{figure}[b]
	\centering
	\includegraphics[width=\linewidth]{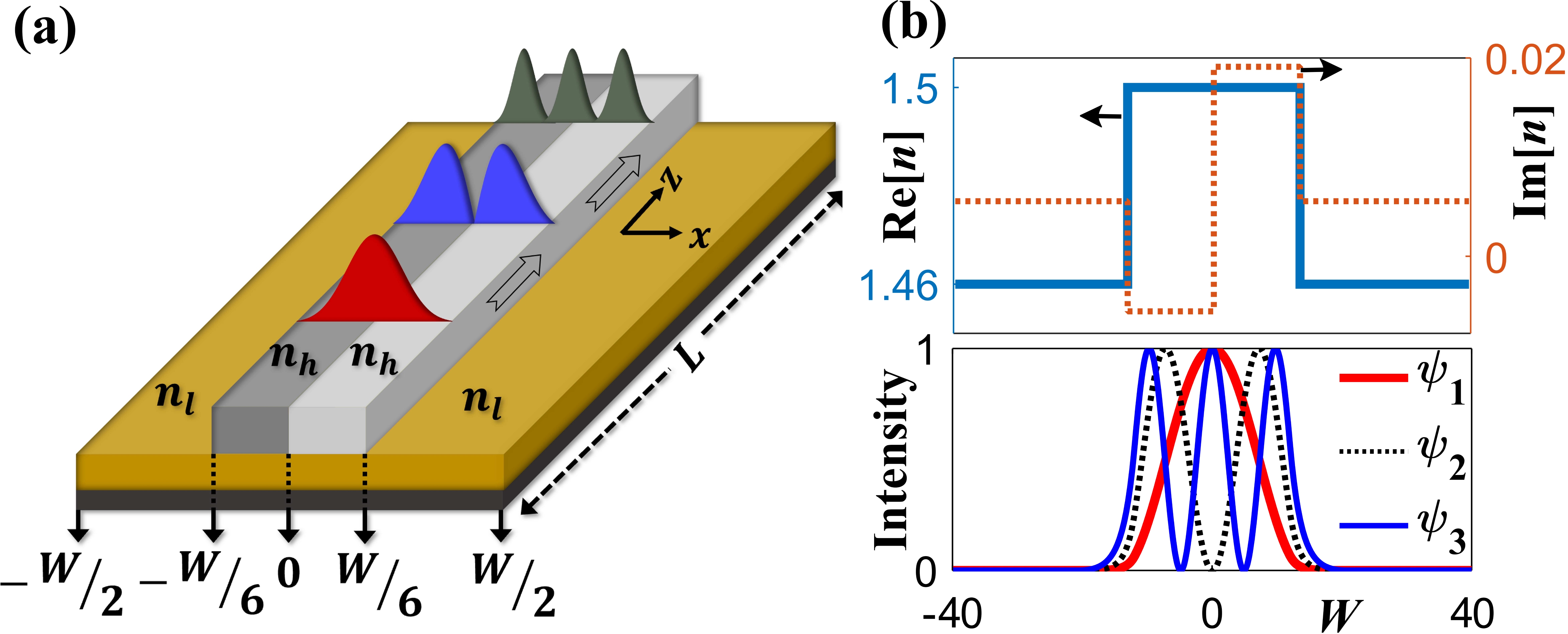}
	\caption{\textbf{Waveguide design}: \textbf{(a)} Schematic of the designed optical waveguide with transverse $x$-axis (considering propagation along $z$-axis). \textbf{(b)} (Upper panel) Transverse refractive index profile $n(x)$ showing $\text{Re}(n)$ (solid blue line) and $\text{Im}(n)$ a specific $\gamma=0.005$ and $\tau=3.45$ (dotted brown line). (Lower panel) Normalized field-intensity profiles of the supported modes $\psi_j\,(j=1,2,3)$.} 
	\label{waveguide}
\end{figure}

To mimic a non-Hermitian system, as shown in Fig. \ref{waveguide}(a), we consider a step-index planar optical waveguide, having a core and a cladding with refractive indices $n_h=1.5$ and $n_l=1.46$, respectively. We normalize the operating frequency $\omega=1$ and set the total width $W=40\lambda/\pi=80$ and operating length $L=15\times10^3$ in a dimensionless unit. To introduce non-Hermiticity in the passive waveguide, we impose a transverse unbalanced gain-loss profile in the following way. 
\begin{equation}
n(x)=\left\{ 
\begin{array}{ll}
\vspace{0.1cm}
n_h-i\gamma, &\quad -W/6\le x\le 0\\
\vspace{0.1cm}
n_h+i\tau\gamma, &\quad 0\le x\le W/6\\
\vspace{0.1cm}
n_l+i\gamma, &\quad W/6\le |x|\le W/2.
\end{array}
\right.
\label{nx} 
\end{equation}
Eq. \ref{nx} represents the overall refractive index profile for a specific cross-section of the waveguide, as shown in the upper panel of Fig. \ref{waveguide}(b); where two independent parameters $\gamma$ and $\tau$ represent gain-coefficient and loss-to-gain ratio, respectively. Obeying Kramers-Kronig causality relation at a single operating frequency \cite{Phang15}, we can independently tune $\gamma$ and $\tau$ along the longitudinal direction to modulate overall non-Hermiticity. For the chosen parameter set, the waveguide hosts three quasi-guided linearly polarized modes, as shown in the lower panel of Fig. \ref{waveguide}(b), that are $\rm{LP_{01}}$, $\rm{LP_{11}}$ and $\rm{LP_{02}}$. In this paper we depict these modes as $\psi_j$ with $j=1,2,3$, respectively, and compute the corresponding propagation constants $\beta_j\,(j=1,2,3)$ using the scalar modal equation $[\partial_x^2+n^2(x)\omega^2-\beta^2]\psi(x)=0$. To control the interactions between the supported modes, in addition with optical gain-loss, we introduce local Kerr-nonlinearity having the form $\Delta n_{\text{NL}}(x,z)=\sigma n_2I$ ($n_2\rightarrow$ nonlinear-coefficient, $I\rightarrow$ signal-intensity and $\sigma=\pm1$ for focusing and de-focusing nonlinearity, respectively); where actual nonlinearity level is quantified in the form of $\left(\Delta n_{\text{NL}}/\Delta n\right)\times100\%$ with $\Delta n=(n_h-n_l)$.
\begin{figure}[t]
	\centering
	\includegraphics[width=\linewidth]{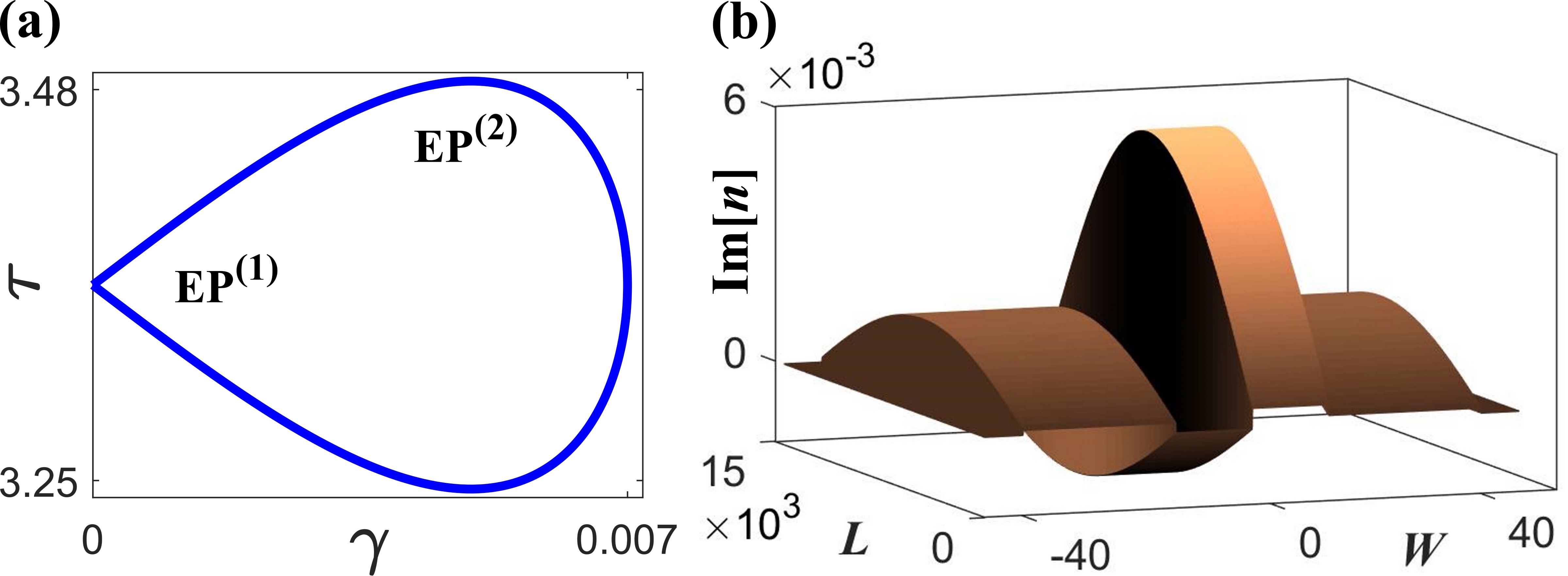}
	\caption{\textbf{Dynamical EP-encirclement}: \textbf{(a)} Chosen topological structure of the waveguide with simultaneous variation of $\gamma$ and $\tau$ around two embedded EPs. Here, EP$^{(2)}$ appears inside the loop only in the presence of nonlinearity. In absence of nonlinearity, the loop encloses only EP$^{(1)}$. \textbf{(b)} Length-dependent variation of Im($n$) after mapping the chosen parameter space as shown in (a).} 
	\label{loss}
\end{figure}

To encounter a second order EP between two coupled modes, we identify the transition between two topologically dissimilar avoided resonance crossing (ARC) phenomena between the $\beta$-values of the corresponding modes with crossing/anticrossing in Re[$\beta$] and Im[$\beta$] in the vicinity of a particular singular point \cite{Laha19,Ghosh16}. Now, in absence of nonlinearity, varying $\gamma$ within the range from 0 to 0.01, we track the dynamics of $\beta_j\,(j=1,2,3)$ for different $\tau$ values and numerically identify an EP at ($\gamma_{EP}=0.0017,\,\tau_{EP}=3.356$) (say, EP$^{(1)}$, as indicated in Fig. \ref{loss}(a)), where $\psi_1$ and $\psi_2$ are analytically connected, however, $\psi_3$ remains unaffected. To encircle the identified EP dynamically, we choose an enclosed parametric loop with a length-dependent distribution of $\gamma$ and $\tau$ following the equations given by
\begin{equation}
\gamma(\phi)=\gamma_{0}\sin\left[\frac{\pi L}{z}\right];\,\,\tau(\phi)=\tau_{EP}+a\sin\left[\frac{2\pi L}{z}\right].
\label{device} 
\end{equation}
Here, $\gamma_0$ and $a$ are two characteristics parameters; where to enclose the EP properly, we have to consider $\gamma_0>\gamma_{EP}$ and $a>0$. The shape of the parametric loop in ($\gamma,\,\tau$)-plane has been shown in Fig. \ref{loss}(a) for a chosen $\gamma_0=0.007$ and $a=0.12$. The corresponding distribution of the Im[$n(x,z)$] has been shown in Fig. \ref{loss}(b). According to the chosen shape of the parameter space in ($\gamma,\tau$)-plane, for both $z=0$ and $z=L$, $\gamma$ must be equal to 0. Thus the complete profile of Im[$n(x,z)$] from $z=0$ to $z=L$ perfectly encloses the EP dynamically, and at the input and output interface, we can get the passive modes, avoiding any loss-dominated modes. Here, one of two different directions of propagation indicates clockwise encirclement and the other indicates anticlockwise encirclement. Here, the propagation of the modes through the waveguide should follow the time dependent Schr{\"o}dinger equation (TDSE) using $z$ as the time axis. Considering paraxial approximation and the variation of Im($n$) within adiabatic limit, we use scalar beam-propagation to solve the equation $2i\omega\partial_z\psi(x,z)=-[\partial_x^2+\Delta n^2(x,z)\omega^2]\psi(x,z)$ (with $\Delta n^2(x,z)\equiv n^2(x,z)-n_l^2$) to study the modal propagations.
\begin{figure}[b]
	\centering
	\includegraphics[width=\linewidth]{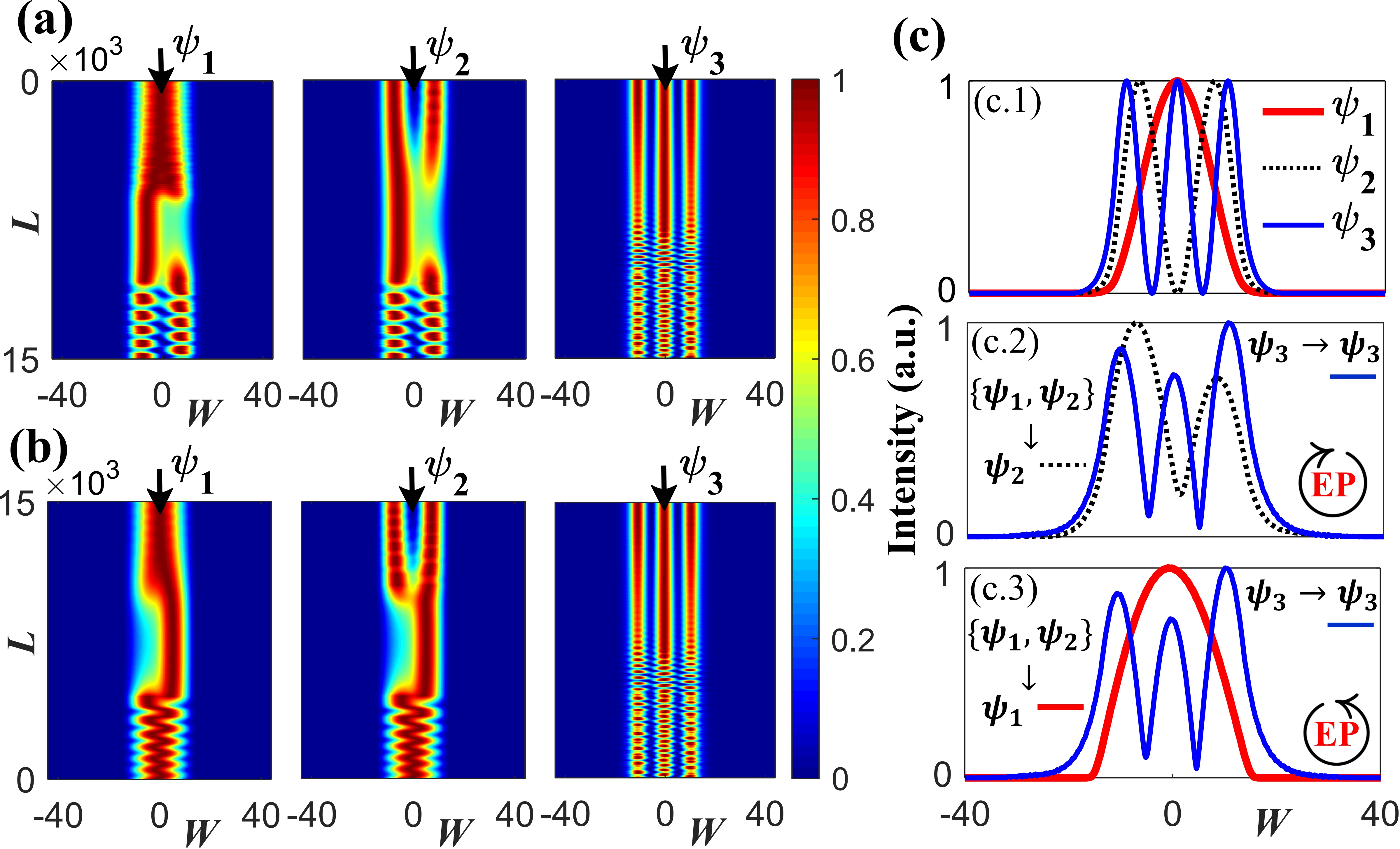}
	\caption{\textbf{Beam propagation simulation results in absence of nonlinearity}: \textbf{(a)} Modal propagations of $\psi_j\,(j=1,2,3)$ with dynamical encirclement of EP$^{(1)}$ in clockwise direction showing adiabatic conversion of $\psi_1\rightarrow\psi_2$ and nonadiabatic evolution of $\psi_2\,(\rightarrow\psi_2)$. \textbf{(b)} Nonadiabatic evolution of $\psi_1\,(\rightarrow\psi_1)$ and adiabatic conversion of $\psi_2\rightarrow\psi_1$, while encircling the EP$^{(1)}$ in anticlockwise direction. For both (a) and (b) $\psi_3$ remains unaffected. \textbf{(c)} (c.1) Supported field intensities. (c.2) Output field intensities at $z=L$ for clockwise EP-encirclement process (considering input at $z=0$). (c.2) Output field intensities at $z=0$ for anticlockwise EP-encirclement process (considering input at $z=L$). We re-normalize the modal intensities at each $z$ for clear visualization and hence the overall intensity variations are essentially scaled.} 
	\label{BP}
\end{figure}

Now, following dynamical encirclement scheme as described in Fig. \ref{loss}, we study the beam propagations of $\psi_j\,(j=1,2,3)$ that have been shown in Fig. \ref{BP}. To implement a clockwise encirclement scheme, the light has been launched at $z=0$ as can be seen in Fig. \ref{BP}(a). Here, both $\psi_1$ and $\psi_2$ associated with EP$^{(1)}$ are essentially converted to $\psi_2$ at $z=L$. Thus, there is one non-adiabatic transition (NAT) corresponding to $\psi_2$. Interestingly, $\psi_3$ is not affected by the presence of EP$^{(1)}$ and retains as $\psi_3$ at $z=L$. Now, launching the light at $z=L$, we implement anticlockwise encirclement scheme; corresponding beam propagation results are shown in Fig. \ref{BP}(b). Here, $\psi_1$ follow a NAT, $\psi_2$ is adiabatically converted to $\psi_1$ at $z=0$, however, $\psi_3$ remains unaffected. In Fig. \ref{BP}(c), we have shown a comparative study among the output field intensities under different launching conditions considered for common excited field intensities at the input (shown in Fig. \ref{BP}(c.1)). While encircling the EP in the clockwise direction, we get the combination of $\psi_2$ and $\psi_3$ at the output; corresponding output field intensities are shown in Fig. \ref{BP}(c.2). On the other hand, while encircling the EP in the anticlockwise direction, the device delivers the combination of $\psi_1$ and $\psi_3$ at the output; corresponding output field intensities are shown in Fig. \ref{BP}(c.3). Thus, owing to the breakdown in adiabaticity during evolutions of the coupled modes, the chiral transmission behavior is evident for dynamical encirclement scheme enclosing only EP$^{(1)}$, even in the presence of non-interacting $\psi_3$. Here, as the parametric loop encloses only EP$^{(1)}$, the overall loss distribution usually affects the corresponding coupled modes $\psi_1$ and $\psi_2$, however, not $\psi_3$; and during transmission, one of the coupled modes that evolves with higher average loss in comparison to other behaves nonadiabatically. For a dual-mode system, an analytical treatment behind such nonadiabatic modal dynamics around an EP has been established in Ref. \cite{Ghosh16,Gilary13}.

Above described investigations have been carried out in the absence of nonlinearity, when $\psi_1$ and $\psi_2$ are interacting with the simultaneous presence of noninteracting $\psi_3$. Now, with the onset of nonlinearity up to 5\%, we further investigate the dynamics of the eigenmodes, where we observe that once we reach 1.2\% nonlinearity, all three modes start interacting mutually. For convenience, we introduce 2.5\% nonlinearity in the spatial index distribution of the waveguide; and studying the mutual interactions and corresponding ARCs between $\beta_j\,(j=1,2,3)$, we encounter multiple second-order EPs. In addition with EP$^{(1)}$ (as described in previous), the waveguide hosts another such a second-order EP in ($\gamma,\tau$)-plane at $\sim(0.0051,3.452)$ (say, EP$^{(2)}$) in the presence of nonlinearity. Here, simultaneous presence of EP$^{(1)}$ and EP$^{(2)}$ analytically connects all the three supported modes $\psi_j\,(j=1,2,3)$. 

\begin{figure}[t]
	\centering
	\includegraphics[width=\linewidth]{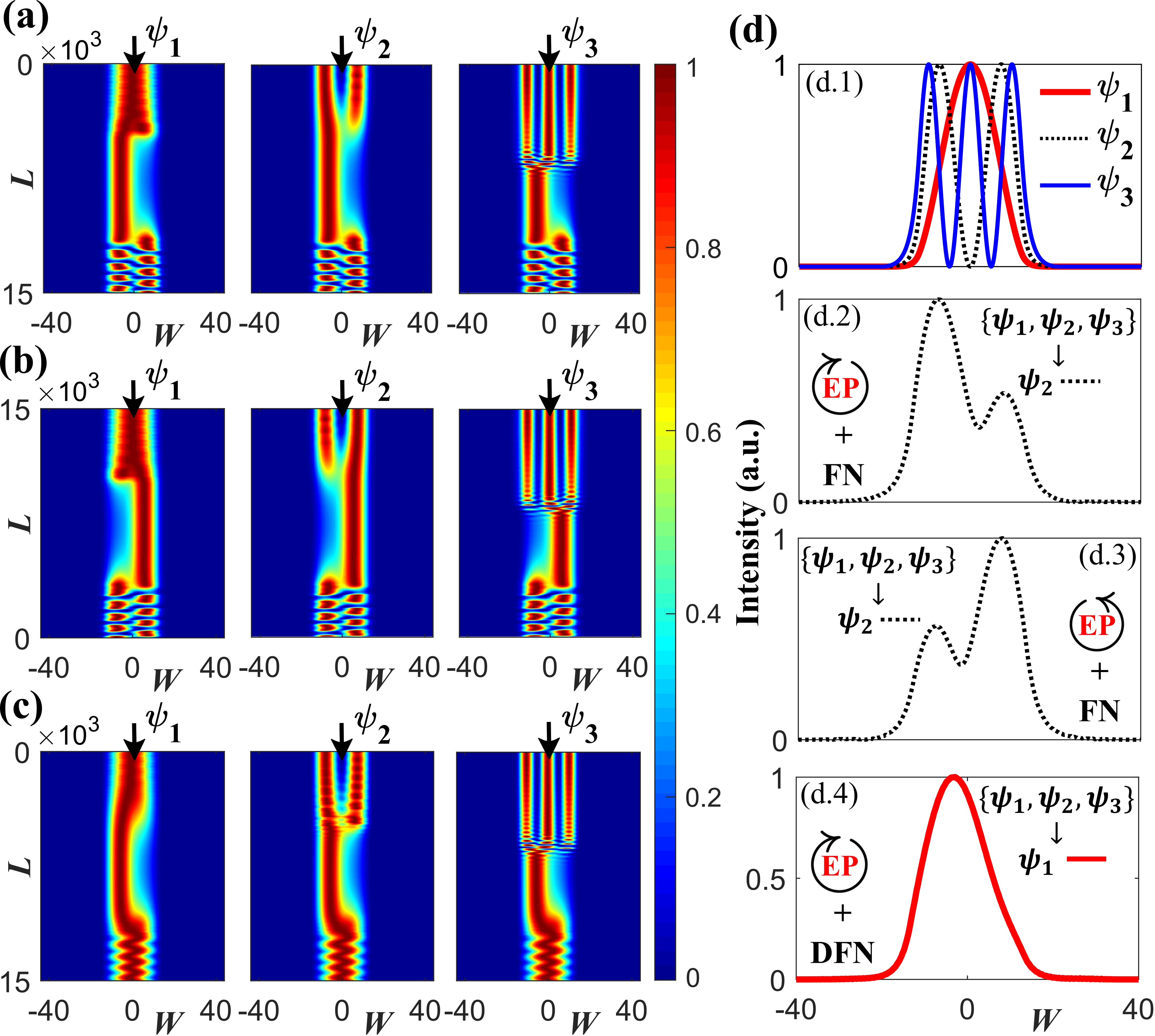}
	\caption{\textbf{Beam propagation simulation results in presence nonlinearity}: \textbf{(a)} In presence of focusing nonlinearity (FN), the propagations of $\psi_j\,(j=1,2,3)$ following dynamical encirclement around two identified EPs (EP$^{(1)}$ and EP$^{(2)}$) in clockwise direction where all the modes are collapsed to $\psi_2$. \textbf{(b)} Similar propagation characteristic of $\psi_j\,(\rightarrow\psi_2)$ showing the nonchiral behavior, while both the EPs have been dynamically encircled in the anticlockwise direction. \textbf{(c)} Propagation characteristic of $\psi_j$ following a clockwise multiple EPs-encirclement process, however, in the presence of de-focusing nonlinearity (DFN) in the optical medium where all the modes are collapsed to $\psi_1$. \textbf{(c)} (c.1) Supported field intensities. In the presence of FN, the output field intensities (c.2) for clockwise and (c.3) anticlockwise encirclement process. (c.4) In the presence of DFN, the output field intensities, while, multiple EPs are encircled in the clockwise direction.} 
	\label{BP_NL}
\end{figure}

We dynamically encircle both EP$^{(1)}$ and EP$^{(2)}$ simultaneously inside the length-dependent parametric loop shown in Fig. \ref{loss}, and perform the beam propagation results in Fig. \ref{BP_NL}. Essentially, we fix the topological structure of the waveguide in such a way that we can consider both the cases, i.e., with and without nonlinearity, given that only in the presence of chosen nonlinearity, $\psi_3$ interacts, but in the absence of nonlinearity, $\psi_3$ behaves as noninteracting state and accordingly EP$^{(2)}$ disappears. Now, we choose $\sigma=+1$ to consider focusing nonlinearity (FN), and considering the light propagation from $z=0$ to $z=L$, we implement a clockwise encirclement scheme in Fig. \ref{BP_NL}(a). Here, this is evident that all the three interacting modes have been collapsed in $\psi_2$, given that $\psi_1$ and $\psi_3$ evolve adiabatically and converted to $\psi_2$, and $\psi_2$ evolves nonadiabatically and retains itself. Now, even we change the direction of light propagation to consider anticlockwise encirclement scheme, we get similar modal dynamics as can be seen in Fig. \ref{BP_NL}(b). Thus, the results, as shown in Fig. \ref{BP_NL}(a) and (b), establish a new nonchiral behavior in modal dynamics for three interacting modes around multiple second-order EPs. Now, in the absence of chirality, we can switch or retrieve a different mode rather than $\psi_2$. Considering $\sigma=-1$, we consider same amount of de-focusing nonlinearity (DFN) and study the dynamics of $\psi_j\,(j=1,2,3)$ in Fig. \ref{BP_NL}(c) for a clockwise encirclement along the same parametric loop described in Fig. \ref{loss}. As can be seen in Fig. \ref{BP_NL}(c), all the coupled modes are collapsed in $\psi_1$ at $z=L$, owing to adiabatic evolutions of $\psi_2$ and $\psi_3$, and nonadiabatic evolution of $\psi_1$. In the presence of de-focusing nonlinearity, the nonchiral behavior in modal dynamics can also be observed by considering the anticlockwise encirclement scheme. In Fig. \ref{BP_NL}(d), a comparative study has been presented where we have shown the output field intensities under different launching conditions considered in the presence of nonlinearity. The commonly excited field intensities (normalized) at the input have been shown in Fig. \ref{BP_NL}(d.1). Figs. \ref{BP_NL}(d.2) and (d.3) show the normalized output field intensities for clockwise and anticlockwise encirclement, respectively, in the presence of FN, whereas Fig. \ref{BP_NL}(d.4) shows the same for clockwise encirclement in the presence of DFN. 

This new and anomalous dynamics of the three interacting modes in presence of multiple EPs can be analytically treated as follows. For intense, consider the $3\times3$ Hamiltonian $\mathcal{H}$ corresponding to the designed waveguide depends on three time dependent parameters $\mu_j(t)$ (for $j=1,2,3$; analogous to $\gamma$, $\tau$ and $\Delta n_{\text{NL}}$). Here, under adiabatic limit, the evaluations of the eigenfunctions of $\mathcal{H}$ follow TDSE. To present a generic mathematics behind nonadiabatic dynamics during conversion between two eigenmodes, we consider the conversions between $\psi^{\text{ad}}_m$ and $\psi^{\text{ad}}_n$ with eigenvalues $\beta^{\text{ad}}_m$ and $\beta^{\text{ad}}_n$; where for our three mode supported waveguide, we can consider $(m,n)\in\{1,2,3\},\,m\ne n$. The corresponding dynamical nonadiabatic correction terms around multiple EPs can be written as 
\begin{subequations}
	\begin{align}
	&\Omega_{m\rightarrow n}^{\text{NA}}=\vartheta_{m\rightarrow n}\exp\left\{-i\oint_0^T\Delta\beta_{m,n}^{\text{ad}}[\mu_j(t)]dt\right\},	\label{omega1} \\
	&\Omega_{n\rightarrow m}^{\text{NA}}=\vartheta_{n\rightarrow m}\exp\left\{+i\oint_0^T\Delta\beta_{m,n}^{\text{ad}}[\mu_j(t)]dt\right\}; 	\label{omega2} 
	\end{align}
	\label{omega} 
\end{subequations}
with
\begin{subequations}
	\begin{align}
	&\vartheta_{m\rightarrow n}=\left\langle\psi_m^{\text{ad}}[\mu_j(t)]\left|\sum_{j=1}^{3}\dot{\mu_j}\frac{\partial}{\partial\mu_j}\right|\psi_n^{\text{ad}}[\mu_j(t)]\right\rangle,\label{omega_p1}\\
	&\vartheta_{n\rightarrow m}=\left\langle\psi_n^{\text{ad}}[\mu_j(t)]\left|\sum_{j=1}^{3}\dot{\mu_j}\frac{\partial}{\partial\mu_j}\right|\psi_m^{\text{ad}}[\mu_j(t)]\right\rangle,\label{omega_p2}\\
	&\text{and}\,\,\Delta\beta_{m,n}^{\text{ad}}[\mu_j(t)]=\beta_m^{\text{ad}}[\mu_j(t)]-\beta_n^{\text{ad}}[\mu_j(t)]\notag\\
	&\qquad\qquad\equiv\text{Re}[\Delta\beta_{m,n}^{\text{ad}}[\mu_j(t)]]-i\Delta\gamma_{m,n}^{\text{ad}}[\mu_j(t)]\label{beta}.
	\end{align}
	\label{omega_p} 
\end{subequations} 
In Eqs. \ref{omega} and \ref{omega_p}, the suffixes $m\rightarrow n$ and $n\rightarrow m$ indicate the conversions $\left|\psi_m^{\text{ad}}\right\rangle\rightarrow\left|\psi_n^{\text{ad}}\right\rangle$ and vice-versa, respectively. In Eq. \ref{omega}, $T$ is the EP-encirclement duration and in Eq. \ref{beta}, $\left|\Delta\gamma_{m,n}^{\text{ad}}\right|$ represents the relative gain between two considering modes. Now, if we consider a situation $\Delta\gamma_{m,n}^{\text{ad}}>0$, then $T\rightarrow\infty$ yields $\Omega_{m\rightarrow n}^{\text{NA}}\rightarrow 0$ and $\Omega_{n\rightarrow m}^{\text{NA}}\rightarrow \infty$. As the pre-exponent terms in Eqs. \ref{omega1} and \ref{omega2} contain the time derivative of three potential parameters, i.e., $\dot{\mu_j}$ as given in Eqs. \ref{omega_p1} and \ref{omega_p2}, the exponential divergence in $T$ of the exponent term of $\Omega_{n\rightarrow m}^{\text{NA}}$ beats the $T^{-1}$ suppression associated with $\vartheta_{n\rightarrow m}$. Thus, for a slow parametric evolution around the EP with in the adiabatic limit, between two considered eigenmodes, only one of them having lower decay-rate evolves adiabatically and the other one behaves non-adiabatically; i.e., at the present condition $\left|\psi_m^{\text{ad}}\right\rangle$ evolutes adiabatically and converted to $\left|\psi_n^{\text{ad}}\right\rangle$, whereas $\left|\psi_n^{\text{ad}}\right\rangle$ follows NAT. 

Now, with proper choices of $m$ and $n$, we can study the possible adiabatic and nonadiabatic conversions between three interacting modes $\psi_j\,(j=1,2,3)$ in our designed waveguide. While we individually consider two different type of nonlinearity ($\sigma=\pm1$) in the optical medium of the waveguide, the modified refractive index profile changes order of $\beta$-values of the supported modes depending on the type of nonlinearity and accordingly the signs of relative gain $\Delta\gamma_{m,n}^{\text{ad}}$ for different choices of $m$ and $n$ are modified. Here, in presence of focusing nonlinearity ($\sigma=+1$), we obtain $\Delta\gamma_{1,2}^{\text{ad}}>0$ and $\Delta\gamma_{3,2}^{\text{ad}}>0$; which gives the adiabatic conversions $\left|\psi_1\right\rangle$ and $\left|\psi_3\right\rangle$ to $\left|\psi_2\right\rangle$, and NAT of $\left|\psi_2\right\rangle$, as can be seen in Figs. \ref{BP_NL}(a) and (b). On the other hand, we obtain $\Delta\gamma_{2,1}^{\text{ad}}>0$ and $\Delta\gamma_{3,1}^{\text{ad}}>0$, while we consider de-focusing nonlinearity ($\sigma=-1$). This yields the conversions $\{\psi_1,\psi_2,\psi_3\}\rightarrow\psi_1$, as can be seen in Fig. \ref{BP_NL}(c), where $\psi_2$ and $\psi_3$ evolve adiabatically, and $\psi_1$ behaves nonadiabatically.

The overall performance of our designed waveguide under different conditions considered throughout this work has been summarized in Table \ref{performance}.
\begin{table}[htpb]
	\caption{Overall device performance}
	\centering
	\begin{ruledtabular}
		\begin{tabular}{ccccc}
			Starting  & Nonlinearity & \multicolumn{2}{c}{End states} & Dynamics \\ \cline{3-4} 
			states    & type $(\sigma)$ & Clockwise       & Anti-        & type   \\ 
			          &              &                 & clockwise    &         \\  \hline
			\cellcolor{gray!20}$\left(\psi_1+\right.$ & $0$ & $\psi_2+\psi_3$ & $\psi_1+\psi_3$ & Chiral  \\ \cline{2-5}   
			\cellcolor{gray!20}$\psi_2+$  & $+1$ & $\psi_2$ & $\psi_2$ & Nonchiral  \\  \cline{2-5}   
			\cellcolor{gray!20}$\left.\psi_3\right)$   & $-1$ & $\psi_1$ & $\psi_1$ & Nonchiral  \\
		\end{tabular}
	\end{ruledtabular}
	\label{performance}
\end{table}

In summary, an exclusive topologically robust and compact nonlinearity induced anomalous mode collapsing phenomenon in a few-mode/multi-mode system has been proposed using the framework of a three-mode supported planar gain-loss assisted optical waveguide that does not bear the chiral property in the presence of multiple second-order EPs. The topological structure of the waveguide in terms of an unbalanced gain-loss distribution is configured in such a way that in the absence of nonlinearity, only two modes are mutually coupled around a single second-order EP keeping the third mode unaffected, whereas, in the presence of nonlinearity, all the three modes are mutually coupled and exhibit multiple second-order EPs. We have been observed that in the absence of nonlinearity, the waveguide exhibits the chiral property even in the presence of a noninteracting mode. Here, depending on the direction of encirclement, only one of two coupled modes survive with the simultaneous presence of third noninteracting mode. Now, with the onset of nonlinearity, we observe that irrespective of the choice of inputs, all the three coupled modes are collapsed in a specific dominating mode, and where the different directions of encirclement around multiple EPs are not able to change the nature of the output due to ruination in the chiral property. Here, we have established that even in the absence of chirality, the individual presence of focusing and de-focusing nonlinearity having the same amount results in different dominating output for the same parametric encirclement process around two EPs. The proposed scheme should also be applicable for systems having more than three states. As irrespective of the choices of the propagation directions, light is converted in a specific state based on the types of the nonlinearities, we may also explore this proposed scheme to achieve nonreciprocal light transmission in a multi-mode system. In the presence of nonlinearity, the proposed new physical aspect of light manipulation around multiple second-order EPs in a multi-state system will certainly provide opportunities in chip-scale integrated photonic devices for next-generation communication systems. 
 
SD acknowledges the support from from Ministry of Human Resource Development (MHRD), India. AL and SG acknowledge the financial support from the Science and Engineering research Board (SERB), Ministry of Science and Technology, India under Early Career Research Scheme [Grant No. ECR/2017/000491].

\end{document}